# Conclusions about properties of high-energy cosmic-rays drawn with limited recourse to hadronic models

A A Watson[*]

*School of Physics and Astronomy, University of Leeds, Leeds LS2 9JT, UK*

**Abstract**

Determining the energy and mass of the highest energy cosmic rays requires knowledge of features of particle interactions at energies beyond those reached at the LHC. Inadequacies of the model predictions set against a variety of data are summarised and it is clear that firm statements about primary mass are premature. Nonetheless, conclusions of significance about the origin of the highest-energy cosmic rays can be deduced from the data.

This paper is dedicated to my great friend and colleague, Jim Cronin, who died suddenly on 25 August 2016, without whom the Auger Collaboration would not have happened.

**1. Introduction**

The origin of the highest-energy cosmic rays, which are explored exclusively through study of the extensive air-shower, remains one of the major puzzles in high-energy astrophysics. One reason is that the majority of the particles are charged so that intervening magnetic fields make it difficult to track them to their birthplace. A second issue is that interpretation of data bearing on mass composition is hampered by lack of knowledge of key features of hadronic interactions, including cross-section, multiplicity, inelasticity and features of pion-nucleus collisions at energies above about 300 GeV, the latter being extremely numerous in air-showers. A collision of a 100 PeV proton with a nucleus has a centre-of-mass energy of ~14 TeV so that the properties of the energy domain beyond this, at trans-LHC energies, are unknown. A conservative assumption is that key properties change smoothly as the energy increases, but one cannot exclude surprises – the *'unknown unknowns'*. Even at ~0.5 PeV, where the air-shower regime begins, the lack of knowledge of some details of the hadronic physics is a serious handicap to extracting mass information and to making accurate estimates of primary energy.

Below I will show first that the extant shower models lead to contradictory deductions about the primary mass at a given energy. Ideally one might envisage that, with the range of independent evidences available, progress in defining the key features of the hadronic interactions would be possible. Although we are still some way from this goal, understanding is increasing. I will then argue that, despite these limitations, key astrophysical information can be extracted from the observations.

**2. Measurements targeted at obtaining the primary mass: evidence for deficiencies in hadronic models**

Above 100 PeV, several methods have been developed that target measurement of the primary mass. The best-known one makes use of fluorescence detectors to study the change of the depth of shower maximum, $X_{max}$, with energy. If, for example, one had a beam of cosmic rays of a single nuclear species then elementary considerations about shower development lead to the expectation that the position at which the particle number, or rate of energy deposition, maximises will move deeper into the atmosphere as the primary energy increases. The rate of change of $X_{max}$ with energy is called the elongation rate, a term introduced by Linsley [1]. Specifically he showed that while for a photon the elongation rate is $2.72 X_0$ g cm$^{-2}$ per decade, where $X_0$ is the radiation length, for a single nuclear species it must be considerably smaller.

---

[*] a.a.watson@leeds.ac.uk

Measurements of $X_{max}$ as a function of energy made at the Auger Observatory [2] are shown in figure 1 together with similar results reported by the Telescope Array (TA) Collaboration [3]. Different approaches to the analyses make point-by-point comparisons impossible. In figure 1 predictions from the Sibyll 2.1 model are also shown.

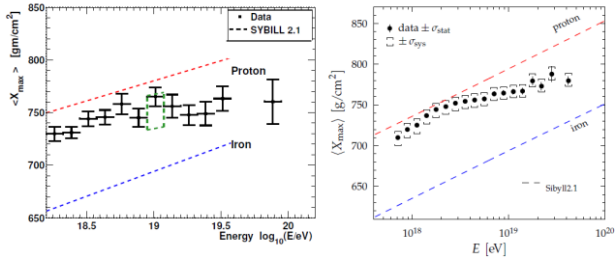

Fig.1. The variation of the depth of shower maximum with energy, as measured by the TA Collaboration (LH) and the Auger Observatory (RH) [3, 2] compared with predictions with the Sibyll 2.1 model for proton and iron primaries.

Neither data set can be fitted adequately with the single straight line expected for a mass composition that does not change with energy: even with the large uncertainties of the TA measurements, the reduced $\chi^2$ for a linear fit to all data is unacceptably large (7.1 for 10 degrees of freedom). Thus the mass composition must be changing as the energy increases, *unless* features of the hadronic model change in a perverse way (for example a marked change in the multiplicity or the cross-section with energy). It is evident that *if* the Sibyll model is correct then both data sets indicate a mass composition that is proton-dominated below ~5 EeV, while at higher energies, because the elongation rate is flatter – as evident in both data sets - the mean mass must be becoming heavier. Note that with a revised Sibyll model it is found that depths of shower maxima are pushed deeper into the atmosphere than found with Sibyll 2.1 [R Engel, these Proceedings].

Moreover, as will be shown shortly, other hadronic models fail to describe data where muons are involved and thus the strongest remark that can be made, independent of model assumptions, is that the mean mass increases above ~5 EeV. This important and unambiguous conclusion is counter to that strongly espoused by the HiRes and TA groups, and often accepted uncritically by theorists, namely that the ultra-high energy cosmic rays are all protons.

If one accepts the models then deductions about the natural logarithm of the atomic mass (ln A = 0 for protons and 4 for Fe) can be made. Estimates of ln A from Auger data using two other models are shown in figure 2 [2]: the Sibyll result lies between those shown. Also displayed are estimates of <ln A> from two other studies. In one [4] the depth of the maximum of the muons in the shower, $<X_\mu^{max}>$, is measured and results compared with predictions. In the other [5], the attenuation of showers across the surface detectors of the Auger Observatory was studied. An asymmetry in the distribution of arrival times of particles is found which is dependent on the development of showers so that mass information can be evaluated. The quantity derived, $((\sec \theta)_{max})$, depends on the radial distance from the shower axis, with muons becoming an increasingly dominant component at larger distances.

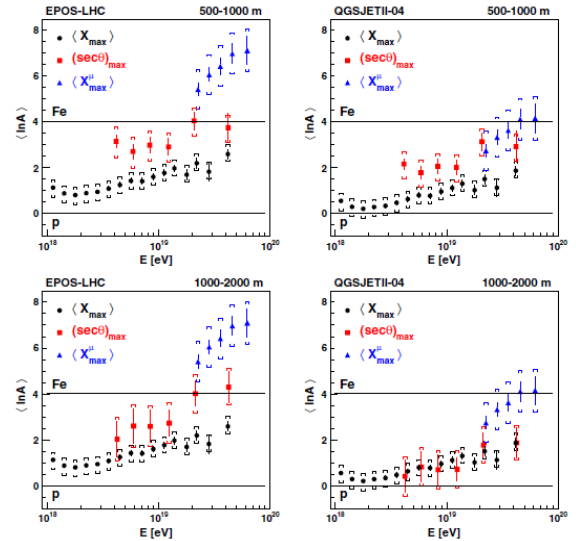

Fig.2. <ln A> as is function of energy as reported in [2, 4, 5] is compared with predictions as a function of energy made using the EPOS-LHC and QGSJETII-04 models. The figure is from [5].

From figure 2 it is evident that the two shower models do not adequately describe the data: an accurate model would be expected to give consistent estimates of <ln A> for different measurements. Further evidences of problems come from sources other than the Pierre Auger Observatory, such as IceTop/IceCube [6], DELPHI [7] and ALEPH [8]. At the South Pole, the IceCube/IceTop Collaboration study muons detected in IceCube in coincidence with showers seen with IceTop. Primary energies from ~4 PeV to ~1 EeV are explored with multi-TeV muons. Estimates of the mean mass as a function of energy from the South Pole work are shown in figure 3 together with those derived from the Auger fluorescence measurements (figure 1, [2]) which are dominated by the electromagnetic component.

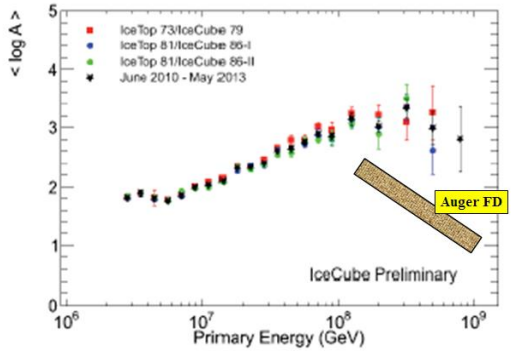

Fig. 3. A comparison of the predictions of the mean value of ln A from South Pole measurements [6] and Auger data [2].

Studies of muon bundles in CERN detectors have been made by the ALEPH [8], DELPHI [7] and, more recently, the ALICE [9] groups at depths of 140 m (70 GeV threshold), 100 m (50 GeV) and 28 m (16 GeV) respectively. The accelerator detectors do not have shower arrays above them. In the earlier work [7, 8] observation of high multiplicity events could not be explained with the hadronic models then available. With ALICE [9], however, it appears that, using the QGSjetII-04 model, the high-multiplicity events can be accounted for. It will be interesting to re-visit these data in future years with revised models.

The Auger Observatory has the unique capability of detecting large numbers of events at high zenith angles (even beyond $80^0$). This enables searches for neutrinos (see below) and the study of showers that are almost exclusively muonic. At $70^0$ the atmospheric thickness traversed before reaching detectors at the depth of the Observatory (875 g cm$^{-2}$) is 2558 g cm$^{-2}$ so that the electromagnetic component produced by $\pi^0$s has ranged-out, leaving a beam of muons (accompanied by an electromagnetic component from muon bremsstrahlung, knock-on electrons, and muon decay that can readily be accounted for). Although the particle distribution on the ground is highly elliptical because of the geomagnetic field, it is possible to measure the number of muons in the shower as a function of energy [10]. Discrepancies are found both in the change in the number of muons with energy and in the number of muons as a function of shower depth (figure 4).

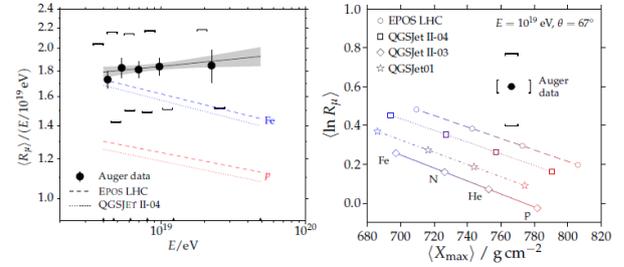

Fig. 4. (LH-figure): average muon content, <$R_\mu$>, as a function of energy. Statistical and systematic uncertainties (square brackets) are shown. (RH-figure): average logarithmic muon content, <ln $R_\mu$>, as a function of shower depth, [10].

Further evidence of problems with current hadronic generators comes from a comparison of events in which the lateral distribution and the longitudinal development have been measured simultaneously. Between 6 and 16 EeV it is found that the average hadronic shower is 1.33 ± 0.16 (1.61 ± 0.21) times larger than is predicted using the EPOS-LHC (QGSJet-04) models with a corresponding excess of muons [11].

Recent fixed-target experiments at CERN [12] may have shed some light on the muon problem. In studies of p-C collisions at 100 and 350 GeV, it is found that the production of the $\rho^0$ meson is larger by ~2 than the production of $\pi^0$s. This is important as the $\rho^0$ decays to two charged pions. It will be some time, however, before the relative production rates of $\rho^0$ and $\pi^0$ are known over a sufficiently large energy range to make model adjustments with confidence.

### 3: Astrophysical conclusions independent of precise hadronic models

There are clearly limitations to the current shower models but this does not prevent some astrophysical conclusions being drawn. In addition to those about the mass of cosmic rays discussed above, information on the arrival direction distribution, the energy spectrum and the neutrino flux can be extracted that is of significant interest.

### 3.1: Arrival Directions

The arrival directions of high-energy cosmic rays have been studied for over 70 years with the expectation that above some energy anisotropies might appear.

There have been many false dawns reflecting the difficulty of the experiments and, sometimes, the over-enthusiasm of the experimentalists. A recent result that is both convincing and important is from the Auger Collaboration who have analysed over 70,000 events above 4 EeV, where triggering of the array is fully efficient for showers out to zenith angles of $80^0$ [13]. Energy estimates at these high energies depend relatively little on hadronic physics (see below).

Above 8 EeV (figure 5) the amplitude of the first harmonic is $(6.4 \pm 1.0)$ % with a probability of this arising by chance of $6 \times 10^{-5}$. With additional data [14, 15], dipole amplitudes above 8 EeV and 10 EeV have been measured as $(7.3 \pm 1.5)$ % and $(6.5 \pm 1.9)$ % with probabilities that these arise by chance of $6.4 \times 10^{-5}$ and $5 \times 10^{-3}$ respectively. The directions of the dipoles are $\alpha = 95 \pm 13^0$ and $93 \pm 24^0$ and $\delta = -39 \pm 13^0$ and $-46 \pm 18^0$. The origin of such dipoles has been discussed by Harari et al. [16a, b] who claim that they could result from diffusive propagation in turbulent extragalactic magnetic fields if the amplitude of the field is large and/or the cosmic rays have a component with high mass. Alternatively the anisotropy could reflect the distribution of the sources which may be similar to that of matter in the local universe.

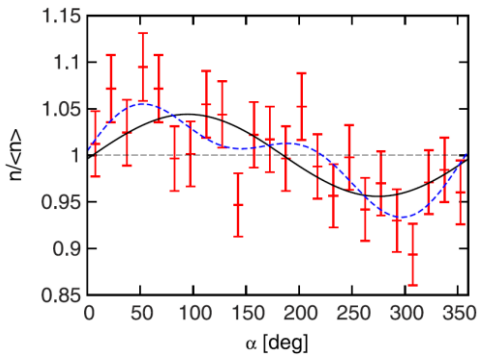

Fig. 5. Observed number of events over the mean as a function of right ascension: 1 sigma uncertainties are shown. The black line shows the first harmonic and the blue line is a combination of the first and second harmonics [13]

At the very highest energies TA have reported a 'hot spot' above 57 EeV in the direction, $\alpha = 148.4^0$, $\delta = 44.5^0$ which is $17^0$ off the super-galactic plane [17], while above a similar energy (58 EeV) the Auger Collaboration find an excess in a region within $15^0$ of Cen A, the closest powerful radio galaxy, with a post-trials significance of 1.4% [18].

The results on arrival directions are independent of assumptions about hadronic physics, though it would be useful if techniques could be devised to separate showers, even crudely, into those developing high in the atmosphere to those developing later. Methods are being explored using features of signals from the surface detectors and from radio emission.

**3.2: The energy spectrum**

One of the earliest features to be identified in the cosmic-ray spectrum was the steepening seen around 3 PeV. The first hints of this feature came in the mid-1950s from measurements of the number spectrum. For decades arguments raged as to whether this was caused by a feature of hadronic interactions or whether it was a reflection of a steepening of the energy spectrum. The matter was resolved, very elegantly, by the KASCADE group who exploited their ability to measure the muon and electron content of showers accurately above ~0.4 PeV. Their results, for two muon energy-thresholds, are shown in figure 6 where the flux of showers that are electron rich (those produced by low mass primaries) are compared with the spectrum of all showers and that of showers that are poor in electrons (heavy primaries) [19].

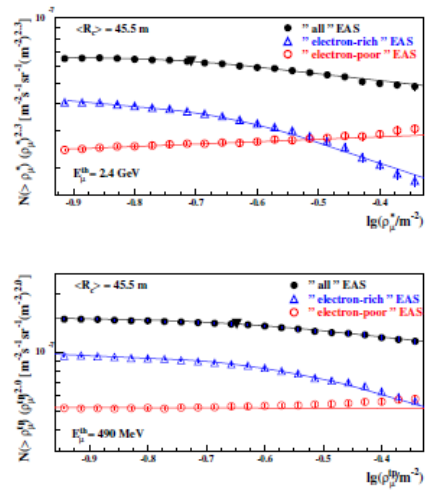

Fig. 6: The number spectra of muons from KASCADE [19]

This beautiful technique has also been applied to data from the KASCADE-Grande project: the results are shown in figure 7 [20a, b].

What can be deduced from figures 6 and 7 is that the flux of the light component of the cosmic rays starts to fall at an energy just above ~3 PeV while the flux of the heaviest component decreases similarly at about 100

PeV, sometimes called the 'second knee'. By contrast, the light component of the spectrum begins to show a harder spectrum above ~ 120 PeV. These conclusions are only weakly dependent on the choice of models. Introducing models one can quantify the break features.

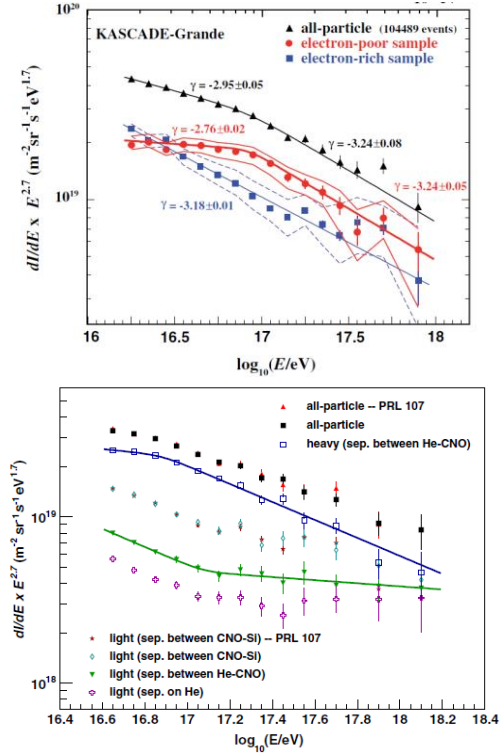

Fig.7: [LH] The reconstructed energy spectrum of the electron-poor and electron-rich components together with the all-particle spectrum o for $0 < \theta < 40^0$ [20a]. The all-particle and electron-rich spectra from the analysis [20a] in comparison to the results of the additional analysis in [20b] with a larger number of events.

Data such as those associated with figure 6 lead to energy estimates for the knee of the energy at the knee to be $4.0 \pm 0.8$ PeV and $5.7 \pm 1.6$ PeV for the QGSJet01 and Sibyll 2.1 models respectively. This gives a strong warning about the dependence of spectral and mass details on models. Similarly, different results on the mass spectra come from the analysis made by the KASCADE group [21], although there is general agreement that there is a knee in the helium spectrum at an energy twice that of the knee in the proton spectrum. The second knee is at an energy ~26 times greater than the knee associated with protons. These results strongly suggest that the spectral features are rigidity dependent and are indicative of the acceleration process and/or the escape processes in the galaxy.

At higher energies the long-standing issue of the shape of the spectrum has now been resolved by the HiRes, Auger and TA groups [22, 23, 24]. The current status is shown in figure 8 [25]. The results of [23, 24] are, to differing extents, largely independent of details of hadronic models. The trick is to use events in which both surface detector and fluorescence detector data are available to construct a calibration curve which can then be used to infer the energy of events that occur during the day or on moonlit nights. A small correction to the energy estimate of ~10% must be made to allow for invisible energy carried into the ground by neutrinos and muons: the correction reduces as the energy increases. Perhaps surprisingly, energy estimates near 10 EeV are much more robust with respect to model uncertainties than those made 4 decades lower.

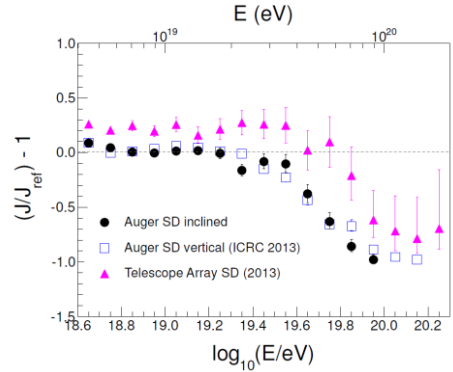

Fig. 8. Fractional differences between the energy spectra of cosmic rays derived by the Auger Collaboration from inclined showers and those from more vertical events from the Auger Observatory and the Telescope Array. The lower energy is the one above which the Auger detector is fully efficient for inclined events [25].

The TA and Auger data show evidence for an ankle in the spectrum at an energy of ~ 5.2 and 4.8 EeV respectively and are in relatively good agreement up to ~16 EeV, with the TA fluxes, on average, 23% greater than those from the Auger Collaboration. However at higher energies there is divergence: the TA fluxes are considerably larger. There could be an interesting astrophysical explanation associated with the TA Hot Spot but, before accepting this, it is necessary to understand issues that arise from the different fluorescence yields used, from the model-dependence introduced in the TA method of calibration and from the mass-dependent correction for invisible energy. This has largely been overcome in the Auger analysis [25, 26] but remains in the TA approach. A joint TA/Auger working group is attempting to resolve these issues using data from the part of the sky common to the two Observatories.

### 3.3 The search for neutrinos

The 1.2 m depth of the Auger water-Cherenkov detectors make searches for high-energy neutrinos possible [27, 28]. The target is to find for showers arriving at large zenith angles that have the characteristics of events from the vertical direction. Such showers are identified from the characteristics of the time distribution of the particles at the water-Cherenkov detectors. No neutrino candidates have been found: the limits (figure 9) are close to those set by IceCube (without the extrapolation and systematic uncertainty in the energy involved in that analysis). The limit is well-below the Waxman-Bahcall bound and just above predictions for proton primaries.

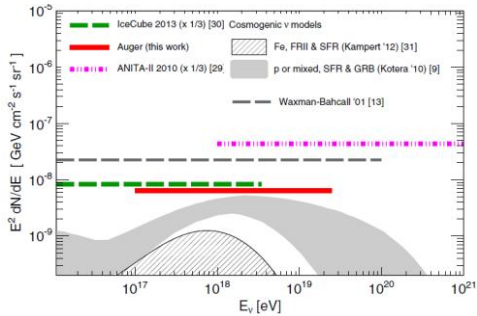

Fig. 9: Upper limt (at 90% confidence level) to the diffuse flux of UHE neutrinos from the Auger Observatory. The predictions of models that assume protons, a mixed composition or heavier primary nuclei are shown [28].

### 3.4 Comment on hadronic uncertainties and predictions of gamma-ray fluxes

Recently evidence of a Pevatron accelerator of cosmic rays has been reported from the Galactic Centre [29]. At such energies there will be systematic uncertainties in the flux of photons arising from the same problems with the underlying hadronic physics that lead to the differences in the estimates of the energy of the first knee of the cosmic-spectrum (section 3.2). Appropriate systematic uncertainties should be shown when γ-ray spectra are reported.

### 4 Tests of some astrophysical models

I will now discuss a few of the models that have been proposed to explain the experimental data. However, one should always keep in mind von Neumann's remark that "with four parameters I can fit an elephant and with five I can make him waggle his trunk".

### 4.1 Testing the Dip model

For many years Berezinsky and his colleagues have argued that the ankle in the cosmic ray spectrum (~5 ± 0.2 EeV) arises from the pair-production of electrons through the interactions of protons with the 2.7 K radiation [30]. Two tests, both weakly dependent on models, have recently been made. In one, an analysis of Auger data has been carried out [31] that shows that the correlation between the depth of shower maximum and the muon content of showers (where the signal at 1000 m from the shower axis is used as a surrogate for direct muon measurements) is inconsistent with proton primaries in the energy range 3-10 EeV. Specifically, compositions with 80% protons and 20% helium, or a 50/50 proton/iron mix, are both strongly excluded for the EPOS-LHC, Sibyll 2.1 and QJSJetII-04 models.

In the second [32] the assumption is made that the Ultra High Energy Cosmic Rays (UHECRs) are protons with the TA spectrum. The values of the source evolution parameter, m, the slope of the spectrum at production, γ, and the maximum proton energy reached in the source are explored. The best fit values of m = 4.3 + 0.4/-0.1, γ = 1.52 +0.35/-0.20 and $E_{max}$ = (50 +5/-1) EeV, are then used to show that the number of neutrinos predicted to be seen by IceCube, if the protons are primaries, is substantially higher than observed (by more than a factor of 3 at the 68% confidence level), a result confirmed independently through the Auger UHE neutrino limit [28] (which might have been cited in [32]). The authors conclude that "an obvious interpretation is that the composition of cosmic rays is heavier than protons at the highest energies, which the Auger composition measurements indicate".

### 4.2 Two models discussing an extra-galactic origin of UHECRs

It is probably the majority view that the highest-energy cosmic rays are of extragalactic origin with debate focussed on the energy above which the extragalactic component dominates. Recently two groups [33, 34] have addressed this question in an interesting fashion. Both assume the sources to be embedded in strong photon fields and calculate the spectrum of particles produced, taking into account the photodisintegration of heavy nuclei and making assumptions about the source spectrum. In [33] GRBs are adopted as the sources while the analysis of [34] is for a more generic situation. I find it rather satisfying that each model provides a natural explanation, through neutron decay, for the dominance of protons near 1 EeV and, because of the extragalactic distribution of the sources, can account for the low level of anisotropy at this energy. The results from the two models are compared with the data in figure [10]

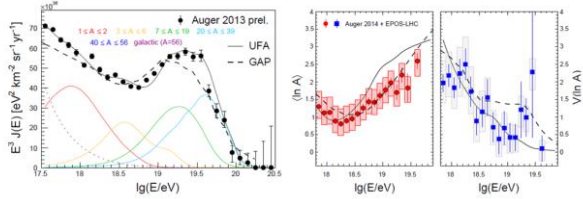

Fig. 10. The extragalactic cosmic ray fluxes as a function of energy for protons, helium and a range of different nuclei. GAP refers to the work of [33] while UFA relates to [34]. The spectrum in the LH plot is from the Auger Collaboration. In the right-hand plots predictions for the evolution of $X_{max}$ and the variance of ln A as a function of energy are compared with Auger measurements [2].

### 4.3 Two models discussing a galactic origin of UHECRs

Some authors continue to argue for a galactic origin of UHECRs. Calvez et al. [35] have proposed that the sources are short gamma-ray bursts (GRBs) occurring in the galaxy at a rate of ~ 1 per $10^5$ years. If the GRBs distributed as the stars in the galaxy, they are able to account for the spectrum with a 90% proton/10% iron mix, but predict an anisotropy that is rather smaller than now reported by the Auger Collaboration. There are also problems with the direction of the anisotropy.

Eichler and colleagues [36, 37] have similarly addressed this possibility. In [36] they conclude that the flux at energies below the ankle arises from sources distributed in proportion to star formation. They claim, for 2.4 EeV, to be able to explain the low anisotropy if the flux is due entirely to intermittent sources. The difficulty with the low anisotropy and the direction reported by the Auger Collaboration [13] is accounted for in the second paper by drift of UHECRs produced beyond the solar circle along the current sheet. Whether the details of composition and spectrum available, plus the significant dipole anisotropy observed, can be accounted for within this model needs further study.

### 4.4: Inferences from the UHECR spectrum and diffuse gamma-ray observations by Fermi

UHECRs interact with photon backgrounds as they travel through space. Some of the energy in UHECRs is expected to cascade down and contribute to the diffuse gamma-ray background. Recently the Fermi LAT Collaboration have shown that above 50 GeV, $86^{+16}_{-14}\%$ of the total extragalactic gamma-ray background can be accounted for by a single population of sources, dominantly blazars [38]. This result has prompted discussion as to whether protons can supply the rest of the background [39, 40, 41, 42]. It could be that there is no room for any diffuse background thus killing the notion that UHECRs are protons, unless the sources are nearby.

In [39] it is suggested that to avoid over-production of the diffuse gamma-ray flux from cascades, there may be a local 'fog' of UHECRs from nearby sources, with the contribution from our galaxy being non-negligible. In this analysis the authors assume that the flux of cosmic rays from 1 – 4 EeV is entirely protonic. In [41] Berezinsky et al argue that this view is too extreme and claim that they can reconcile the Extra Galactic Radiation Background (EGRB) flux with proton primaries. This view is supported by a rather similar independent analysis by Supanitsky [42]. In [40] it is shown that only sources that evolve like BL Lacs would produce sufficiently low levels of secondary radiation and thus remain as a possible source of UHECRs.

There is thus a fascinating connection between the EGRB below about 1 TeV and UHECRs. It is possible that this line of study, along with the limits set by the lack of observation of high-energy neutrinos, will be important in answering questions about the origin of the UHECRs and in particular about their mass, a crucial question for future projects to search for neutrinos.

### 5: Summary and conclusions:

It is clearly disappointing, after so many years of experimental effort, that even the age-old question of galactic vs extragalactic origin of the highest-energy cosmic rays cannot be answered with complete certainty. A major limitation of the measurements is that neither spectra nor anisotropies can yet be studied as a function of the mass of the particles with adequate statistical precision. While the fluorescence technique yields precise measurements, the data are too sparse to make mass-dependent anisotropy searches. However, methods are being developed to use the information from the surface detectors and the radio technique has promise for the measurement of the depth of shower maximum in large samples of events.

In this paper I have argued that, despite the limitations in our knowledge of the hadronic physics, firm statements can be made relating to the nature and origin of high-energy cosmic-rays. These are:

1. The 'knee' in the cosmic ray spectrum at ~3 PeV is due to the loss of protons and other light nuclei because of a rigidity cut-off in the accelerator and/or associated with escape from the galaxy.

2. There is clear evidence of a second knee near 100 PeV which is associated with a diminution in the flux of heavy nuclei such as iron. Again this is an accelerator and/or an escape effect.

3. Near 1 EeV, the mass spectrum is rich in light nuclei, but we cannot be sure what fraction is protonic because of the uncertainty in the hadronic physics.

4. There is a softening of the cosmic-ray spectrum near 5 EeV, the 'ankle'.

5. It is beyond doubt that the energy spectrum steepens above about 30 EeV.

6. The paucity of high-energy neutrinos seen by the Auger and IceCube Observatories argues against the bulk of the highest-energy cosmic rays being protons.

7. The Fermi-LAT data present interesting challenges to the idea that the bulk of the highest energy particles are protons.

8. The anisotropy seen found in Auger data above 8 EeV will be difficult to account for in terms of a galactic origin.

And what of the future? The Telescope Array is being expanded by a factor of 4 to address the specific question of the reality of the hot-spot [D Ikeda, these Proceedings]. At the Auger Observatory, 4 m$^2$ of scintillator are being put above each water-Cherenkov detector to aid the identification of muons and thus help discover what fraction of the highest energy cosmic rays are protons and also to shed light on the muon puzzle [G Cataldi, these Proceedings]. The upgrades at both Observatories will be completed in the next few years. The Stratospheric Super-Pressure balloon flight planned for a EUSO-style detector, planned for 2017, is a key step towards the location of such a device on the International Space Station [P Gorodetsky, these Proceedings]. The results from these ground-based, balloon and space endeavours will dictate the future of the field.

**Acknowledgements:** I am very grateful to the organisers for inviting me to such an enjoyable meeting. I also take this opportunity to thank my Auger colleagues for many discussions over the last 25 years but the views expressed in this paper do not necessarily reflect those of any of them. Remarks by Ralph Engel and Michael Unger have been particularly helpful and Michael's provision of figures 1 and 10 is gratefully acknowledged. Space limitations have prevented me from discussing many interesting models of cosmic ray origin.